\documentclass[traditabstract]{aa}

\usepackage{txfonts}
\usepackage{natbib}
\usepackage{graphicx}
\usepackage{times}
\bibpunct{(}{)}{;}{a}{}{,}
\newcommand{\ea}{\hbox{$\epsilon$~Aur }}
\newcommand{\eax}{\hbox{$\epsilon$~Aur}}
\newcommand{\eae}{\hbox{$\epsilon$~Aurigae }}

\newcommand{\p}{$\pm$ }

\newcommand{\m}{$^{\rm m}\!\!.$}

\newcommand{\kms}{km~s$^{-1}$ }
\newcommand{\kmsx}{km~s$^{-1}$}

\newcommand{\teff}{$T_{\rm eff}$ }
\newcommand{\logg}{{\rm log}~$g$ }
\newcommand{\ms}{M$_{\odot}$}
\newcommand{\rs}{R$_{\odot}$}
\newcommand{\ANG}{\accent'27A}
\newcommand{\Am}{\ANG~mm$^{-1}$ }

\newcommand{\ha}{H$\alpha$ }
\newcommand{\hax}{H$\alpha$}
\newcommand{\hb}{H$\beta$ }

\newcommand{\fe}{\ion{Fe}{ii} }

\newcommand{\si}{\ion{Si}{ii} }

\newcommand{\ubv}{\hbox{$U\!B{}V$}}
\newcommand{\bv}{\hbox{$B\!-\!V$}}
\newcommand{\ub}{\hbox{$U\!-\!B$}}

\begin{document}

\title{Spectral and photometric analysis of the eclipsing binary \eae
prior to and during the 2009--2011 eclipse
\thanks{Based on spectra obtained at the Dominion Astrophysical Observatory,
Ond\v{r}ejov Observatory and Castanet-Tolosan Observatory and on
\ubv\ photometry gathered at the Hvar Observatory and Hopkins
Phoenix Observatory.}
\thanks{Tables 1 and 2 are available only in electronic form
 at the CDS via anonymous ftp to {\tt cdarc.u-strasbg.fr} (130.79.128.5)
 or via {\tt http://cdsweb.u-strasbg.fr/cgi-bin/qcat?J/A+A/.}}
}
\titlerunning{Spectral and photometric analysis of \eae
prior to and during the 2009--2011 eclipse}

\author{P.~Chadima\inst{1}\and P.~Harmanec\inst{1}\and
P.~D.~Bennett\inst{2,3}\and B.~Kloppenborg\inst{4}\and
R.~Stencel\inst{4}\and S.~Yang\inst{5}\and
H.~Bo\v{z}i\'c\inst{6}\and M.~\v{S}lechta\inst{7}\and
L.~Kotkov\'a\inst{7}\and M.~Wolf\inst{1}\and
P.~\v{S}koda\inst{7}\and  V.~Votruba\inst{7}\and
J.L.~Hopkins\inst{8}\and C.~Buil\inst{9}\and D.~Sudar\inst{6}}

\institute{Astronomical Institute of the Charles University, Faculty
of Mathematics and Physics,\\ V~Hole\v sovi\v ck\'ach~2,
CZ--180~00~Praha~8, Czech Republic \and Eureka Scientific, Inc.,
2452 Delmer Street, Suite 100, Oakland, CA 94602-3017, USA  \and
Department of Astronomy \& Physics, Saint Mary's University,
Halifax, NS B3H 3C3, Canada \and Department of Physics and Astronomy,
University of Denver, 2112 East Wesley Avenue, Denver, Colorado 80208,
USA \and Department of Physics and Astronomy, University of Victoria,
P.O. Box 3055 STN CSC, Victoria, B.C., Canada V8W 3P6 \and Hvar Observatory,
Faculty of Geodesy, Ka\v{c}i\'ceva~26, 10000~Zagreb, Croatia \and
Astronomical Institute of the Academy of Sciences, CZ--251 65
Ond\v{r}ejov, Czech Republic \and Hopkins Phoenix Observatory, 7812
West Clayton Drive, Phoenix, Arizona 85033, USA \and Castanet
Tolosan Observatory, 6 Place Clemence Isaure, 31320 Castanet
Tolosan, France}

\offprints{P.~Chadima,\\
\email pavel.chadima@gmail.com}
\date{Received February 17, 2011; accepted April 5, 2011}

\abstract{A series of 353 red electronic spectra (from three observatories,
mostly from 6300 to 6700~\ANG) obtained between 1994 and 2010, and 
of 171 \ubv\ photometric observations (from two observatories)
of the 2010 eclipse, were analyzed in an effort to better understand
\eax, the well-known, but still enigmatic eclipsing binary with the 
longest known orbital period ($\sim$27 yrs). The main results follow.
(1) We attempted to recover a spectrum of the companion by disentangling 
the observed spectra of the \ea binary failed, but we were able to 
disentangle the spectrum of telluric lines and obtain a mean spectrum of 
the F-type primary star. The latter was then compared to a grid of
synthetic spectra for a number of plausible values of \teff and \logg$\!\!$,
but a reasonably good match was not found. However, we conclude that 
the observed spectrum is that of a low gravity star. (2) We examined 
changes in the complex \ha line profiles over the past 16 years, with 
particular emphasis on the 2009--2011 eclipse period, by subtracting a
mean out-of-eclipse \ha profile (appropriately shifted in radial velocity)
from the observed spectra. We find that the dark disk  around the unseen
companion has an extended \textquoteleft atmosphere\textquoteright\  that
manifests itself via blueshifted and redshifted \ha \textquoteleft shell\textquoteright\ 
absorptions seen projected against the F star. Significantly, the \ha shell
line first appeared {\sl three years before first contact of the optical
eclipse} when the system was not far past maximum separation. (3) Analyses
of radial velocities and central intensities of several strong, unblended 
spectral lines, as well as \ubv\ photometry, demonstrated that these
observables showed apparent multiperiodic variability during eclipse.
The dominant period of 66\fd21 was common to all the observables, but
with different phase shifts between these variables. This result strongly
supports our earlier suggestion that the photometric variability seen
during eclipse is intrinsic to the F star, and therefore, the idea 
of a central brightening due to a hole in the disk should be abandoned.  
Although variability on similar timescales is also seen in the spectrum
and in photometry out of eclipse, we were unable to find a coherent 
periodicity in these data. Nevertheless, these results appear to rule
out regular stellar pulsations as the cause of this variability.

}
\maketitle

\section{Introduction}\label{intro}
Epsilon Aurigae (7 Aur, HD 31964, HR 1605) is a bright star in the
constellation of Auriga ($V\sim$3$^{\rm m}\!\!.$0, r.a. $05^{\rm h}01^{\rm
m}58^{\rm s}$, decl. +43$^\circ$\rm49$^\prime$\rm24$^{\prime\prime}$).
It is a single-line eclipsing binary with the longest known orbital
period of 27.1 years (9890\fd3) \citep{Ludendorff1903, Chadima2010}.
Even more notable than the orbital period is the long duration of
the eclipse of almost two years, which in combination with the large
binary separation means that an eclipsing object must be huge.
Analyzing the primary eclipse, \citet{Huang1965} concluded that
the eclipsing object was a cool opaque disk. No secondary eclipse
has ever been observed. An object in the center of the disk is
invisible, and various hypotheses about its true nature had been put
forward, since there is no easy way to derive the individual
masses and radii. It is therefore conceivable that \ea belongs to
those relatively rare binaries (like $\beta$~Lyr) for which the
brighter component in the optical spectral region can be the less
massive of the two. To avoid confusion, throughout this paper we
use the terms {\sl primary} and {\sl secondary} to denote the
visible F-type star and the unseen object (presumably hidden in the 
cool disk), respectively.

Two main models of the system have been considered, often referred
to as the high-mass, and the low-mass model \citep[][and references
therein]{Guinan2002}.

{\sl The high-mass model} was proposed in a remarkable study
by \citet{Kuiper1937} and consists of an F0Ia primary of
$\sim$36$M_{\odot}$ and a 1200--1400~K cool secondary hidden
in the opaque disk, with a mass of $\sim$24.5$M_{\odot}$. They also
estimated the radii of 190 and 2690~\rs. The separation of the
binary components is then $\sim$35~AU. They pointed out that the
observed flat-bottomed eclipse can only be understood if the cool
eclipsing body is an extended gaseous shell. The model was later
developed by \citet{Carroll1991} who considered the F0Ia primary
of $\sim$15$M_{\odot}$ and the secondary hidden in the opaque disk,
with a mass of $\sim$13.7$M_{\odot}$. The separation of the binary
components in their model is $\sim$27.6~AU. The disk surrounding
the secondary is considered to have a protoplanetary origin.

{\sl The low-mass model} assumes a bloated post-AGB object and a 
secondary hidden in an optically thick disk, which is assumed to
originate from recent accretion of matter flowing from the F primary
to secondary. This model was first formulated by \citet{eggl85} who
suggested the masses of 1.3$M_{\odot}$ for the F star and 5$M_{\odot}$
for the secondary. Along with \citet{liss84}, they could not rule out
the possibility that the object hidden in the disk is itself a binary, 
thus making \ea a triple system.

Although these two models are very different ones, there were no
firm observational constraints to prefer one over the other. Very
regrettably, the distances that follow from the Hipparcos parallax
\citep[362--2273~pc;][]{esa97} or from the revised reduction
\citep[355--4167~pc;][]{leeuw2007a,leeuw2007b} are so imprecise
that they are consistent with both models. \citet{Hoard2010} assembled
new Spitzer Space Telescope IRAC observations of \ea, and after
combining them with archival ultraviolet (UV) to mid-infrared data,
they obtained a spectral energy distribution (SED) from $\sim$0.1~$\mu$m
to 100~$\mu$m. They argue that this SED can be reproduced by a
three-component model consisting of a 2.2\p0.9$M_{\odot}$ F type
post-AGB star (the primary) and a 5.9\p0.8$M_{\odot}$ B5V star
(the secondary) surrounded by a geometrically thick but partly
transparent disk of gas and dust with an effective temperature
of 550~K. However, we note that the far-UV spectrum of \ea observed
by the FUSE satellite is an emission line spectrum, probably
produced by scattering of continuum photons from a hot star embedded
in the occulting disk of the secondary \citep{Bennett2005, Ake2006}.
Future models of the spectral energy distribution should take this
into account.

Some additional support for the high-mass model comes from the
recent work of \citet{Sadakane2010}, who performed an LTE abundance
analysis of \ea using ATLAS models and applying NLTE corrections.
\citet{Sadakane2010} compared \ea to the reference star HD~81471,
an A7~Iab supergiant. They find that both stars have comparable
\teff and \logg$\!\!$, but that \ea had higher microturbulent velocity.
Overall abundances of the elements Mg, Si, S, Ca, and TI, and of Sc,
Cr, and Fe are comparable and close to solar in both stars. The
$\sl s$-process elements Y, Zr, and Ba are slightly more abundant
(by +0.25 dex) in \ea than in HD~81471, but Sr is anomalously low
in {\sl both} stars. Carbon and oxygen are underabundant, and N and
Na overabundant, in both stars. This abundance pattern is typical
of late-type supergiants and is quite different from that of post-AGB
stars. In particular, for the four examined post-AGB stars,
\citet{Sadakane2010} find enhanced carbon abundances, with
[C/Fe] $>$ [N/Fe] for these stars (in contrast to the situation
for \ea and HD 81471). They conclude that the observed abundances
of \ea are normal for high-mass supergiant stars. 

\begin{figure*}[th]
\centering \resizebox{\hsize}{!} {\includegraphics{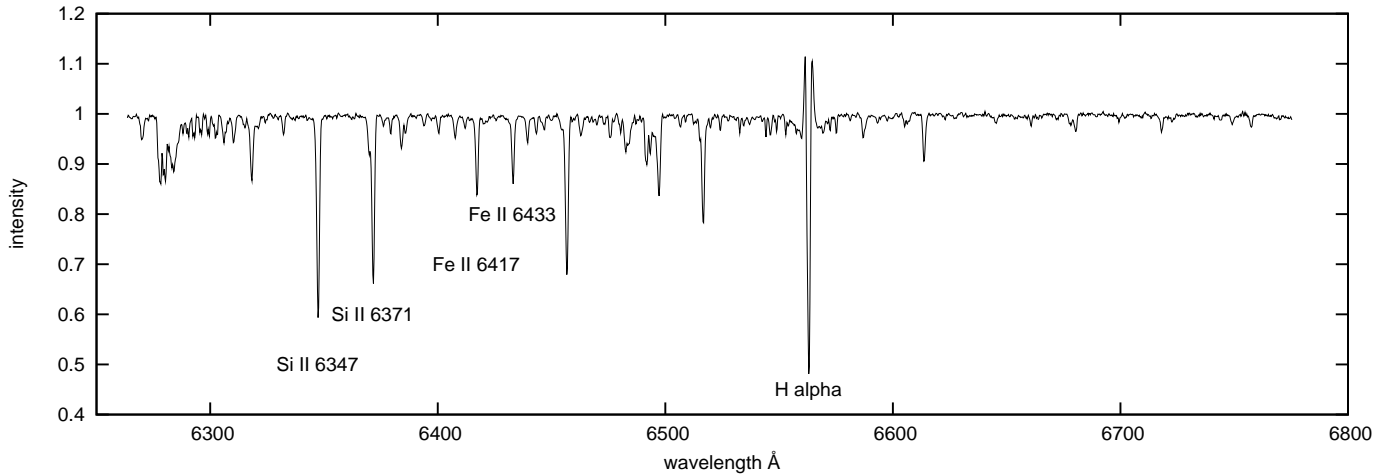}}
\caption{Sample spectrum of \ea from OND, showing the strong, unblended lines 
of \si and \fe for which the radial velocity, central intensity, and
equivalent width were measured.}
\label{spectrum}
\end{figure*}

\begin{figure*}[p]
\centering \resizebox{\hsize}{!}{\includegraphics{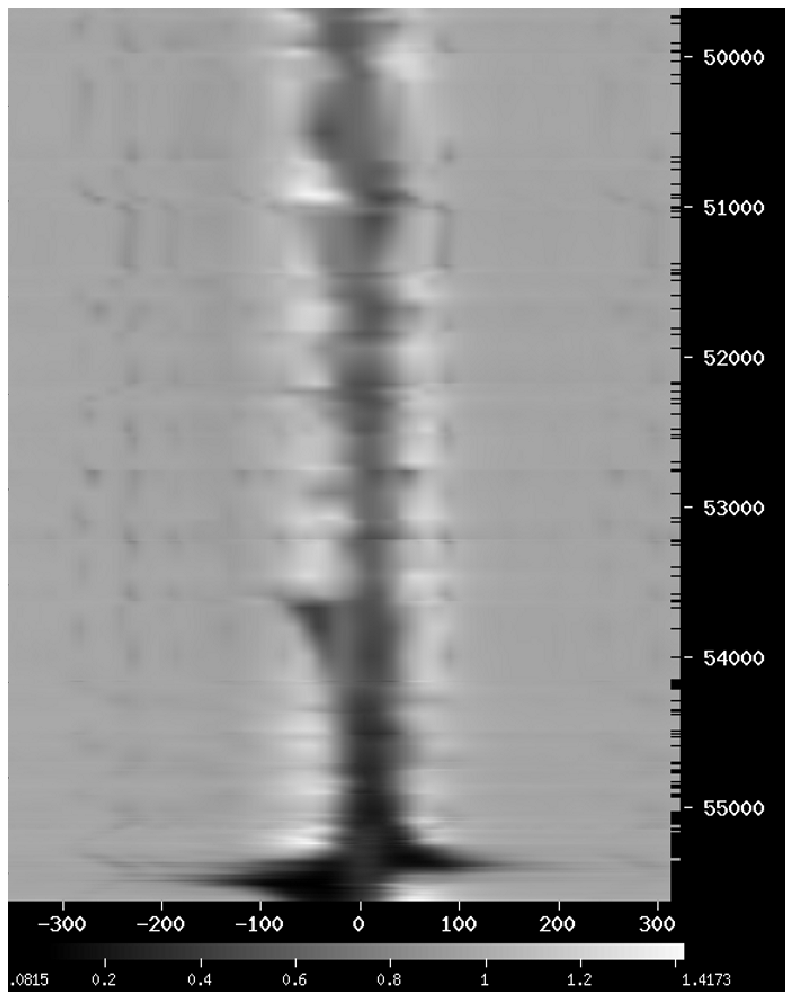}}
\caption{Evolution of the \ha profile from 1994 to present.
The x-axis shows heliocentric velocities (in \kmsx) and the y-axis
shows observation dates (HJD-2,400,000). The intensity scale is shown
below. The spectra were linearly interpolated onto a regular time
grid to produce this image. The black dashes along the right axis
show the actual observation dates.}
\label{ha_fig1}
\end{figure*}

\begin{figure*}[p]
\centering \resizebox{\hsize}{!}{\includegraphics{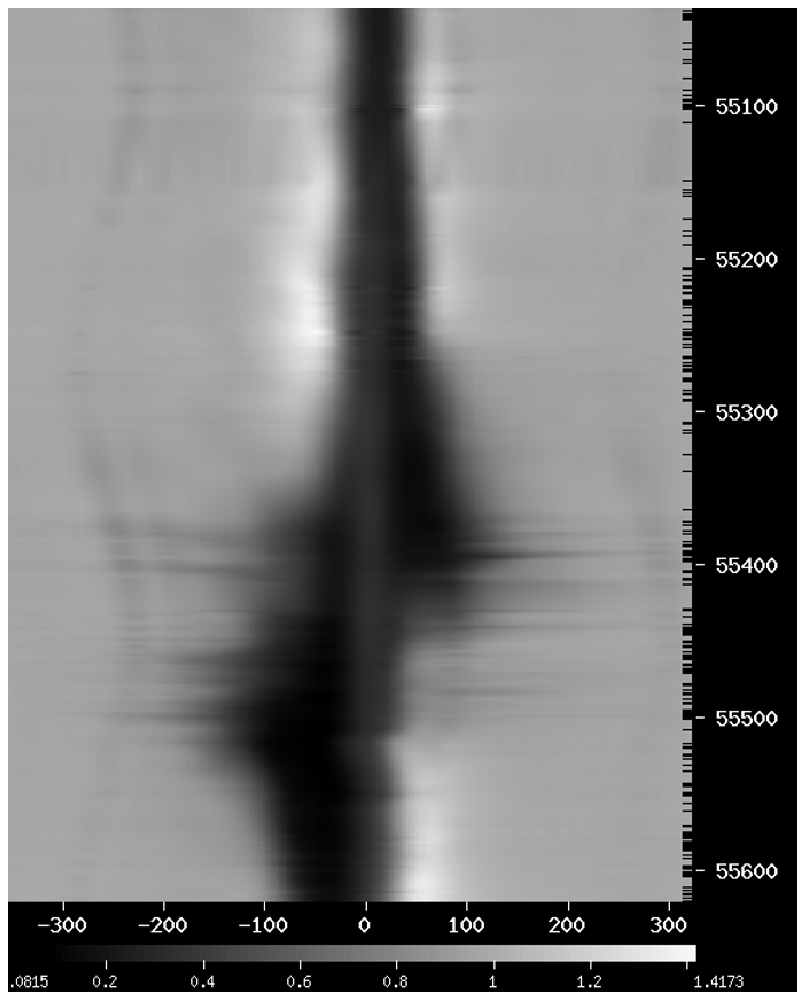}}
\caption{Evolution of the \ha profile during the 2009--2011 eclipse.
The x-axis shows heliocentric velocities (in \kmsx) and the y-axis
shows observation dates (HJD-2,400,000). The intensity scale is shown
below. The spectra were linearly interpolated onto a regular time
grid to produce this image. The black dashes along the right axis
show the actual observation dates.}
\label{ha_fig2}
\end{figure*}

It would be impractical to summarize all of the published studies of \eax.
We restrict ourselves to those relevant to the findings presented in
this paper. Several investigators, starting with \citet{Kuiper1937},
have noticed the presence of a \textquoteleft shell\textquoteright\ 
spectrum during the eclipse, which manifests itself by the presence
of additional absorption lines, obviously originating in the optically
thin outer atmosphere of the disk. This shell spectrum can be used to
analyze of the properties of the disk around the secondary. \citet{Hinkle1987}
observed CO shell lines, which are absent in the spectrum of the primary,
and \citet{Lambert1986} found that the K\,I lines at 7664 and 7699~$\AA$
strengthen noticeably at the beginning and during the eclipse.
Measuring their radial velocity (RV) and assuming a Keplerian
disk, they estimated probable masses of the components of \ea as less than
3~\ms, and 3--6~\ms, for the primary and secondary, respectively, and
argued in favor of the low-mass model. They also observed cyclic line-profile
variations on a timescale of about two months and noticed that two
known pulsating F-G\,Ia supergiants, HR~8752 and $\rho$~Cas, have
much longer periods of about 1 year. They speculated that, if the
two-month variations are indeed caused by photospheric pulsations of
the primary, this may lend additional support to the conclusion that
its mass is low. \citet{Saito1987} analyzed several shell lines and
attempted to derive some properties of the disk. Since the shell
lines were blended with the lines of the primary, they corrected the
eclipse spectra, dividing them by one reference spectrum taken
out-of-eclipse. We note, however, that since the \ea spectra are
intrinsically variable, it may not be optimal to use only one,
accidentally chosen reference spectrum for such a correction. Indeed,
\citet{Ferluga1991} used several out-of-eclipse spectra to derive
an \textquoteleft average\textquoteright\  reference spectrum, which
they used for the reconstruction of 15 pure shell spectra. They then
measured the physical quantities characterizing the properties of a
number of shell lines originating in the disk. \citet{Backman1985} analyzed
hydrogen Brackett lines during the 1982--1984 eclipse. They concluded 
that the emission and absorption velocities of the B$\alpha$ and B$\gamma$
lines have a time dependence more like what is expected for the primary
than the secondary, but with a large uncertainty.

Readers are also referred to a review paper by \citet{Guinan2002}
where detailed pieces of information about this mysterious system, as
well as important references, can be found.

Recently, a long and extensive series of archival radial velocity
beginning in 1899, and photometric observations of eclipses since 1848,
have been compiled and analyzed independently by two groups. The results
were the new, more precise ephemerides and orbital solutions of
\citet{Stefanik2010} and \citet{Chadima2010}. These are clearly superior 
to the previous spectroscopic orbital solution of \citet{Wright1970}
since they are based on much longer observational periods and incorporate 
more accurate RV measurements. Moreover, they simultaneously use
photometric measurements to derive a more accurate orbital period and
mideclipse epoch. The ephemeris of \citet{Stefanik2010} is
\begin{equation}
T_{\rm prim.min.} = {\rm HJD}\,(2455413.8 \pm 4.8) +
(9896^{\rm d}\!\!.0 \pm 1^{\rm d}\!\!.6) \times E,
\end{equation}
and it redicts the current mideclipse somewhere between July~31 and
August~10,~2010. \citet{Chadima2010} derived the ephemeris
\begin{equation}
T_{\rm prim.min.} = {\rm HJD}\, (2455402.8 \pm 1.0)
+ (9890^{\rm d}\!\!.26 \pm 0^{\rm d}\!\!.62) \times E\label{efe},
\end{equation}
which predicts the current mideclipse to occur on July 24--26, 2010. 
This orbital solution is used in this paper. Furthermore, \citet{Chadima2010}
argue that the brightening observed during the middle of the 1982--1984
eclipse was not due to a central hole in the disk, but was a continuation
of the intrinsic F~star variations observed out of eclipse.

The current eclipse began August 2009 and is predicted to end in May 2011.
Naturally, this rare event has attracted the attention of many astronomers
in an attempt to better understand the nature of this still mysterious
system. To this end, the present collaborative effort has obtained a diverse
variety of observations using the best available instruments and observational
techniques and modern analysis methods. An outstanding example of such work
is the recent study by  \citet{Kloppenborg2010} in which the 2-D image of
\ea was reconstructed from CHARA multi-aperture, multi-baseline interferometry.
Preliminary results obtained during the current eclipse show an image 
of the primary star being partly eclipsed by a disk-like companion, 
thereby providing direct confirmation of the \citet{Huang1965} model.
Assuming a mass of 5.9~\ms\ \citep{Hoard2010} for the B star hidden
in the disk, they estimated the mass of the F star as 3.6{\p}0.7~\ms\ .

\citet{Leadbeater2010} report a step-wise increase in the line
strength of the K\,I line  at 7699~$\AA$ during the current eclipse,
which seems to support an earlier suggestion by \citet{fer90} that
the disk has a multiring structure. These and other recent findings
about \ea have been summarized by \citet{Stencel2010}.

The present study is organized as follows. The observational
material and its reduction and measurement are presented in
Sect.~\ref{obs}. Our attempts at disentangling the \ea spectra and
the comparison of disentangled spectra of the primary with the
synthetic spectra are discussed in Sect.~\ref{disen}. Analyses and
possible interpretation of the complex and highly variable \ha line
profiles are given in Sect.~\ref{ha}. An analysis of the measured radial
velocities and central intensities of several spectral lines and
\ubv\ photometry is presented in Sect.~\ref{per}.

\section{Observational material, reduction, and measurements}\label{obs}
\subsection{Spectroscopy}
This study is based on the following three series of electronic
spectra covering the spectral region around the \ha line.
\begin{enumerate}
\item 105 CCD spectra with a linear dispersion of 10~\Am and a two-pixel
resolution of 21700 were secured in the Coud\'e focus of the 1.22-m reflector
at the Dominion Astronomical Observatory (DAO) in Canada between
1994 and 2010 by SY and PDB. These observations cover the wavelength 
range 6200--6750~\ANG.
\item 201 CCD spectra having a linear dispersion of 17~\Am and resolution
of 12600 were obtained in the Coud\'e focus of the 2.0-m reflector
of the Ond\v{r}ejov Observatory (OND) in the Czech Republic 
in 2006--2010 by PH, P\v{S}, M\v{S}, MW, PC, VV, and several
additional observers (credited in Acknowledgements). These observations
cover the wavelength range 6260--6760~\ANG.
\item 47 CCD spectra with resolution of 11000 were obtained with a 0.28-m
Celestron 11 telescope at the Castanet-Tolosan Observatory in France
(CTO hereafter) by CB during 2009--2010.
\end{enumerate}

Initial reductions of the DAO and OND spectra were carried by SY and
M\v{S}, respectively, both in IRAF. The initial reductions of the
CTO spectra were carried out by CB in Reshel V1.11. Rectification
and all radial velocity, central intensity (CI), and equivalent width 
(EW) measurements of selected lines in these spectra were carried out
by PC in SPEFO \citep{Horn1996, Skoda1996}. The latest JK~2.63 version
of the program SPEFO, developed by the late Mr.~J.~Krpata, was used.
All measurements are presented in detail, together with heliocentric
Julian dates (HJDs) of mid-exposures in Table~1 (electronic only).

A representative Ond\v{r}ejov red spectrum is displayed in
Fig.~\ref{spectrum}. Quantitative measurements  were carried
out for several of the stronger, unblended spectral lines:
two \ion{Si}{II} lines at 6347.109~$\AA$ and 6371.371~$\AA$,
two \ion{Fe}{II} lines at 6416.919~$\AA$ and 6432.680~$\AA$,
and the \ha line at 6562.817~$\AA$, for which the emission wings
were also measured. The RV measurements in SPEFO are based on
sliding the direct and reverse line profile on the computer
screen until the best coincidence of both profiles is achieved.
The zero point of the RV scale was determined individually for each
spectrum via measurements of selected telluric lines \citep{Horn1996}.

The behavior of the \ha profile is illustrated well by the 2-D images
of the line profile over time, shown in Fig.~\ref{ha_fig1} for the
entire 1994 to 2011 period, and in Fig.~\ref{ha_fig2} in detail for the
2009--2011 eclipse. Note the high degree of variability of the \ha profile.
The velocity scale for these spectra is heliocentric, and one can see
the annual sinusoidal variations of several nearby telluric water vapor
lines in Fig.~\ref{ha_fig2}. A detailed analysis in the \ha variations
is presented in Sect.~\ref{ha}.

\subsection{Photometry}
Here, we use the following two series of photoelectric \ubv\
observations from the phases of the current total eclipse.
\begin{enumerate}
\item 105 differential observations relative to $\lambda$~Aur (HD~34411)
in a period HJD~55213--55468 were obtained by HB, DS, PH, MW, and PC with
a photoelectric photometer and an uncooled EMI 9789QB tube at the Hvar 
Observatory (Croatia) 0.65-m Cassegrain reflector. Observations in all
three filters consisted of from 7 to 10-second integrations followed
by 5-second integrations on the sky. The check star HR~1644 (HD~32655)
was observed as frequently as the variable and typically three to five
such individual differential observations were obtained for each night of 
observations. All these observations were reduced and transformed to the 
standard \ubv\ system with the help of non-linear transformations using 
the reduction program HEC22 \citep{Harmanec1994, Harmanec1998}. The latest 
release \#17 of the program HEC22 was used. This version allows modeling
the varying extinction that occurs during the course of each night's
observations. The standard \ubv\ magnitudes of the comparison star $\lambda$~Aur

\smallskip
$V=4$\m705, \bv = 0\m621, and \ub = 0\m142
\smallskip

\noindent were derived from numerous all-sky observations at Hvar and
were added to the magnitude differences {\it var-comp} and {\it check-comp}
to obtain final differential \ubv\ magnitudes of \ea and HR~1644. All
Hvar observations of both stars were tabulated with the HJDs in Table~2
(electronic only).
\item 66 differential observations, also obtained relative to
$\lambda$~Aur in a period HJD~55200--55539 were secured by JH at the Hopkins
Phoenix Observatory, USA. His home-built photometer uses a 1P21
photomultiplier tube operated at --950~V and is attached to a
Celestron C-8 telescope mounted on a Meade LX-90 mount. Dead time
for the detector is determined using an aperture mask. The mask
contains irregularly spaced holes that reduce the aperture by 80\%.
Stars that are close to each other and near the zenith are measured
as quickly as possible to reduce the effects of extinction. A
bright star and a faint star are measured without, with, and without
the aperture mask placed in the light path. Their resulting counts
are averaged and used to compute the dead time coefficient of the
instrument. Observations are conducted in the standard CVCVCVC
format (C = comparison, V = variable).  Each observation is composed
of three 10-second integrations in each of the three \ubv\ filters,
followed by one 10-second integration on the sky, again in \ubv. The
following values of the comparison star, slightly different from the
Hvar values, were adopted:

\smallskip
$V=4$\m71, \bv = 0\m63, and \ub = 0\m12.
\smallskip

\noindent These data\footnote{Available from:\\ 
{\tt http://www.hpsoft.com/EAUr09/Data/UBVRIData.html}} 
are on the instrumental \ubv\ system of the instrument.
\end{enumerate}

\begin{figure*}[t]
\centering \resizebox{\hsize}{!}{\includegraphics{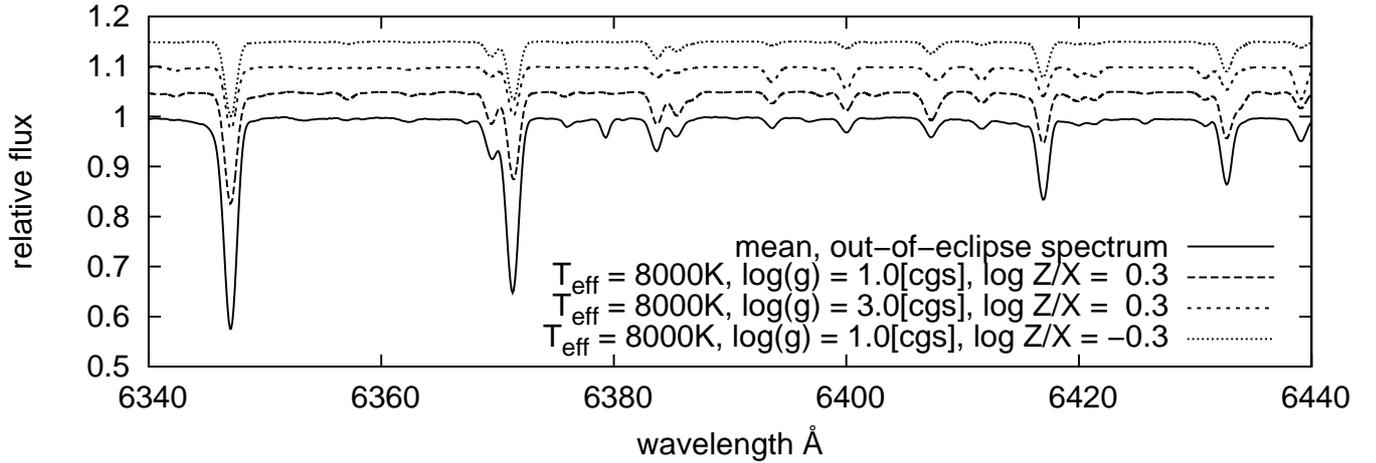}}
\caption{Comparison of the mean, out-of-eclipse spectrum of \ea with 
some ATLAS synthetic spectra. The best-fitting synthetic spectrum
has the parameters: \teff = 8000~K, \logg = 1.0[cgs]
log $Z/X$ = --0.3, $v$(rot) = 5.0~\kmsx. Two other synthetic spectra
are shown, one with higher gravity and one with lower metallicity.}
\label{comparison}
\end{figure*}

\section{Disentangling and modeling the \ea spectra}\label{disen}
Having at our disposal several hundred \ea spectra, which span an
interval of 17 years, i.e. more than a half of the orbital period,
we attempted to disentangle them in the hope of detecting a faint
spectrum of the \textquoteleft unseen\textquoteright\  secondary,
possibly lost in the much more prominent spectrum of the primary.
Note, for instance, that if the true mass ratio of the system was
not too far from one, the lines of both components would remain
blended in all orbital phases since the semi-amplitudes of their
RV curves would both be only $\sim$15~\kms. For the purposes of 
disentangling spectra, we only used observations obtained prior 
to the predicted beginning of the current eclipse. This left us
with 208 spectra (99 DAO and 109 OND spectrograms). We mainly used
the interval 6333--6540~$\AA$ and also several shorter subintervals
(e.g. an interval of 6344--6374~$\AA$, containing both \si lines).
We made no attempt to disentangle \ha, considering its complicated
shape, obvious $V/R$ changes of its emission wings, and the
possibility that it need not be associated with either the primary
or secondary component.

We used the program KOREL, developed by \citet{Hadrava1995,
Hadrava1997, Hadrava2004} and made publicly available by the author.
It uses a spectral disentangling technique in the Fourier domain
and simplex method of minimalization of the sum of residuals
and discrete spectra from various orbital phases, which are
rescaled to a linear RV scale. From these data, the spectra of the
individual binary or multiple components are derived, as well as, if
required, the orbital elements of the binary or multiple system.
No assumption is made about the shape of the line profiles;
the only requirement inherent to the method is that the component
profiles can only vary in their strength and not in their shape.
The run of KOREL is controlled by a parameter file with the initial
orbital parameters of the system and initial simplex steps. Any of the 
orbital parameters can either be held fixed or be converged.

We selected only a short wavelength interval 6529--6539~$\AA$ containing
virtually no stellar lines but only a number of stronger telluric lines,
then fixed the orbit of the Earth around the Sun projected into the
direction to \eax, and let the program derive the relative strength
of the telluric spectrum for each spectrogram used. Then we fixed these
strengths and also all elements of the orbital solution derived by
\citet{Chadima2010}. We instructed the program to proceed step by
step, first allowing for the determination of variable line strengths
for both binary components only and finally also allowing for
convergence of the epoch of periastron, which is the most uncertain
parameter in the adopted orbital solution.

To see whether the resulting sum of squares of residuals $\chi^2$ is
sensitive to the mass ratio $q$ used (thus indicating the definite
presence of a secondary line spectrum), we carried out a number of KOREL
solutions for different values of $q$. Such a mapping has been
successfully applied in several other cases -- see, for
instance, \citet{hec2010}.

We applied this mapping to the whole selected wavelength interval, as
well as to several sub-intervals but the expected, roughly parabolic
dependence of $\chi^2$ on the mass ratio has ever been found. We must
conclude that, at least for the spectral resolution we had at our
disposal, no lines of the secondary spectrum were detected.

After obtaining this null result, we decided to use KOREL only for
disentangling the telluric spectrum and recovering the mean 
out-of-eclipse spectrum of the F-type primary. Using the mean
stellar spectrum minimizes the effect of transient, intrinsic 
variations.

We used a catalog of model atmospheres ATLAS for our calculation
of synthetic spectra. We calculated a grid of the synthetic spectra
of \textquoteleft classical\textquoteright\  F stars based on Kurucz'
ATLAS9 model atmospheres for various temperatures, gravities, and
metallicities. In particular, the grid was derived for \teff = 7500--8500~K,
\logg = 1.0--5.0[cgs], and metallicities log~$Z/X$ from --0.3 to +3.0.
All synthetic spectra were convolved to rotation velocities of the primary
in an interval $v$(rot) = 5.0--25.0~\kmsx. Synthetic spectra for higher
metallicities and lower gravities were not modeled because the ATLAS
catalog does not contain model atmospheres for those parameters owing
convergence problems.

The best match between the disentangled and synthetic spectrum (in the
sense of the lowest value of $\chi^2$) was obtained for a synthetic
spectrum with the following characteristics: \teff = 8000~K, \logg =
1.0[cgs], log~$Z/X$ = --0.3, $v$(rot) = 5.0~\kmsx. Our results are in
fair agreement with the spectral analysis by \citet{Sadakane2010}.
In Fig.~\ref{comparison}, the disentangled spectrum is compared with
several synthetic spectra, and it is clear that even the model for
\logg =1.0 does not describe the observed spectrum properly; that is to
say, the two \si lines (on the left side of the plot) and two \fe lines
(on the right side of the plot) should be more prominent than they
are for the best synthetic spectrum with \logg = 1.0.

To check on the result, we investigated the variation in $\chi^2$ as
a function of the input quantities, which characterize each model.
We also inspected the plots of individual spectra. The conclusion
from this exercise is that the synthetic spectra are almost
insensitive to the value of the projected rotational velocity for
the values used, at least in comparison to other characteristics.
This implies that the \textquoteleft optimal\textquoteright\  value
of the projected rotational velocity should not be given too much weight,
even more so because a satisfactory fit between the disantangled and
model spectra was not found. The dependence on \teff is almost negligible
for the range of investigated values and the spectral lines used.
There is, however, a very strong dependence on the gravity and
metallicity. In Fig.~\ref{comparison}, there are a few synthetic
spectra for lower metallicity and higher gravity. It is seen that
one could probably reproduce the \si and the \fe lines by increasing 
the metallicity and further decreasing \logg. Regrettably, we do not 
have a suitable model atmosphere available to confirm this conjecture.

However, the best-fitting synthetic spectrum already provides a good
fit to the weaker lines from 6375--5415 $\AA$, even though it fails
to adequately describe the stronger \si and \fe lines. This behavior
illustrates the limitations of the ATLAS model atmospheres used here,
because they are plane-parallel LTE models. Both of these assumptions
are probably significantly violated in low-gravity, highly evolved
stars (regardless of whether the F~star is of low or high mass).
Therefore, these ATLAS models should be considered as providing only
{\it qualitative constraints on the stellar parameters}
(\teff, \logg, abundance). For quantitative work, non-LTE (NLTE)
models, and probably spherically symmetric models also, are needed.

\section{Analysis of the \ha profile during the 2009--2011 eclipse}\label{ha}
The profile of the \ha line of \ea observed outside the eclipses
consists of a central absorption surrounded by a double emission
peaks, which might originate in the extended envelope of the primary
\citep{Cha1994}. The whole \ha profile and, especially, its emission
wings are very variable, which is sometimes explained by the
rotation and physical changes in the envelope around the primary.
This variability causes problems in the interpretation of the \ha
profile even out of eclipse.

To investigate the assumption that the \ha emission originates
in a vicinity of the primary, we calculated orbital solutions for
the \si lines and the \ha emission wings using RV measurements on the DAO
out-of-eclipse spectra. We used the FOTEL program \citep{Hadrava2004} with
input parameters from the solution of \citet{Chadima2010} and
only allowed a convergency of the periastron epoch $T$ and the velocity
semi-amplitude $K_1$. Results are summarized in Table~\ref{tab3}.

\begin{figure}[t]
\centering \resizebox{\hsize}{!} {\includegraphics{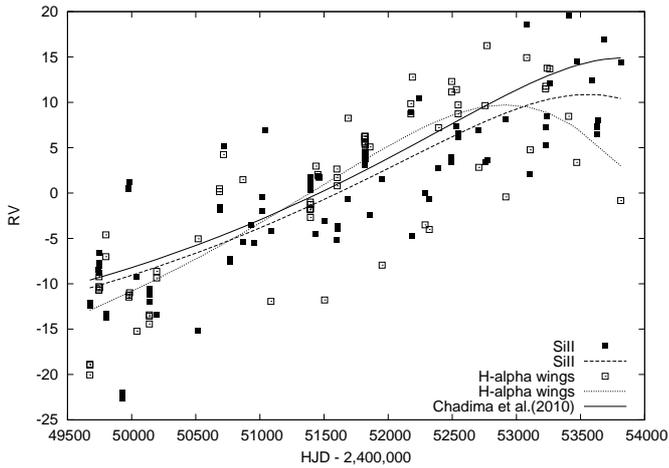}}
\caption{Orbital solutions derived only from the \si lines and from
the emission wings of \ha are compared to the full \citet{Chadima2010} 
solution.}
\label{orbit}
\end{figure}

In Fig.~\ref{orbit}, the orbital solution found from the \si lines
alone and the solution from the wings of \ha are compared with the
solution of \citet{Chadima2010}. It is seen that the \ha emission
roughly follows the orbital motion of the \si lines, i.e., the
primary, and therefore must be associated with this component. We
infer that a small difference between both solutions is mainly
caused by an insufficient interval covered by data used for both
orbital solutions (when compared with the orbital period of the
\eax).

\begin{figure}[t]
\centering \resizebox{\hsize}{!} {\includegraphics{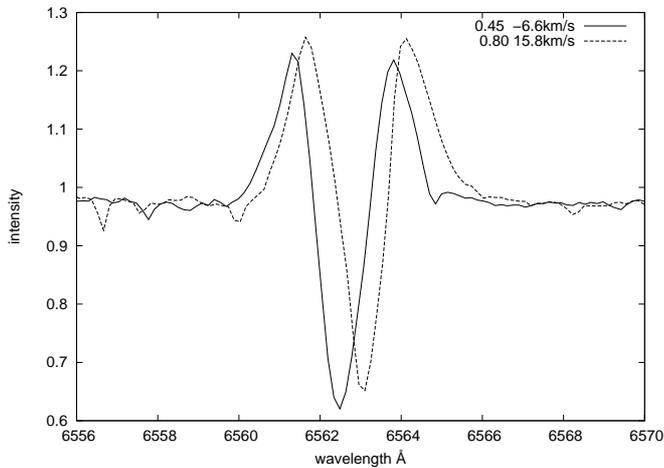}}
\caption{Comparison of two symmetric \ha profiles from different
orbital phases. The orbital phases and the radial velocities
of the primary at those phases are labeled.} \label{halfa}
\end{figure}

To verify that the \ha emission originates near the primary,
we selected two rather symmetric \ha profiles from substantially
different orbital phases and overplotted them on a heliocentric
wavelength scale in Fig.~\ref{halfa}. One can see that the whole
\ha profile, including the double emission wings, undergoes the RV
shift corresponding to the motion of {\sl the F star} in orbit.
This behavior provides strong evidence that the \ha emission is
associated with the F-type primary.

\begin{figure}[t]
\centering \resizebox{\hsize}{!} {\includegraphics{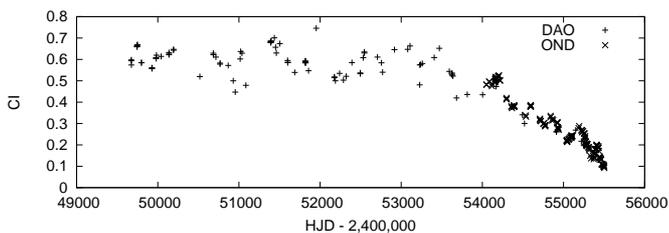}}
\caption{Variation of the \ha core central intensity in the
DAO and OND spectra.}
\label{Ha_orig}
\end{figure}

\setcounter{table}{2}
\begin{table}
\caption{Orbital solutions for $T$, $K_1$, $v_0$ from fits 
to the \si lines and \ha wings.}
\label{tab3}
\begin{center}
\begin{tabular}{lccc}
\hline\hline\noalign{\smallskip}
Spectral line & $T$ & $K_1$ & $v_0$\\
& (HJD) & (\kmsx) & (\kmsx)\\
\noalign{\smallskip}\hline\noalign{\smallskip}
\si       & 2454280\p270 & 13.1\p1.0 & --4.6\p1.6\\
\ha wings & 2453640\p130 & 16.1\p1.1 & --9.3\p1.3\\
\noalign{\smallskip}\hline
\end{tabular}
\end{center}
\end{table}

During eclipse, \ha undergoes prominent line profile changes.
Figure~\ref{Ha_orig} shows a remarkable fact: the CI of the
\ha absorption core steadily decreases with the approaching
eclipse. An important finding is that this decrease {\sl started
about three years before the beginning of the optical eclipse},
thus confirming the behavior first reported by \citet{Kuiper1937}.
The \ha EW increased during the same time period. The RV of
the central absorption core increased significantly during the
first half of the eclipse, but after mideclipse, then suddenly
reverted back to negative values. The \ha emission peaks disappear
during eclipse. For the current eclipse, the red emission peak had
disappeared by HJD~2455259 and the blue peak had vanished by HJD~2455329.
The red emission peak subsequently reappeared after HJD~2455524.

From a study of the \ha line profile during the last eclipse
\citep{Barsony1986}, we expect the evolution of the \ha profile
in the second half of the current eclipse to follow the sequence
of events described above in reverse. The observations should
show a gradual increase in the \ha absorption core CI, accompanied by
a decrease in its EW, and a gradual return of the absorption core RV 
to the out-of-eclipse values. It is expected that both emission
peaks will reappear in reverse order to what was seen in the
first half of the eclipse.

Based on the results of previous studies of the shell spectrum
(discussed in Sect.~\ref{intro}), we adopted the \citet{Kuiper1937}
interpretation of the \ha behavior.  We assume that (apart from any
intrinsic line profile changes, as seen for other lines) changes
in the \ha profile during eclipse are due to additional absorption
of the primary's light by the outer, optically thin atmosphere
of the companion's occulting disk. In other words, we suppose that,
since the companion's disk does not contribute any radiation in
the optical region near \hax, the uneclipsed part of the primary
star produces a similar \ha profile to the whole stellar disk out
of eclipse. However, the region producing the \ha emission may be
extended, in which case much of the emission may not be eclipsed
during the stellar eclipse, thereby also resulting in a similar
eclipse emission profile to the one out of eclipse. We interpret
the disappearance of the out-of-eclipse \ha emission during eclipse
(Fig.~\ref{ha_fig1}) as the result of strong shell absorption at
the corresponding wavelength. In this scenario, the intrinsic
emission is still present, but is strongly absorbed by material
in the occulting disk during eclipse. This disk absorption reflects
the velocity field in the rotating disk, which is why its RV increases
in the first half of the eclipse. This naturally leads to an apparent
disappearance of the red emission peak. As the eclipse goes on,
the shell absorption from the other edge of the disk, rotating in
the direction towards the observer, becomes dominant and the observed
RV of the deepest absorption quickly shifts to negative (i.e.,
blueshifted) values, thus leading to an  apparent disappearance
of the blue emission peak. Near the center of the eclipse, both
edges of the rotating disk atmosphere are projected against the
primary and one should simultaneously see a blueshifted and a
redshifted absorption. If the disk were axially symmetric, one
should see two absorption cores of the same CI with opposite RVs
with respect to the instantaneous RV of the secondary.

\begin{figure}[t]
\centering \resizebox{\hsize}{!} {\includegraphics{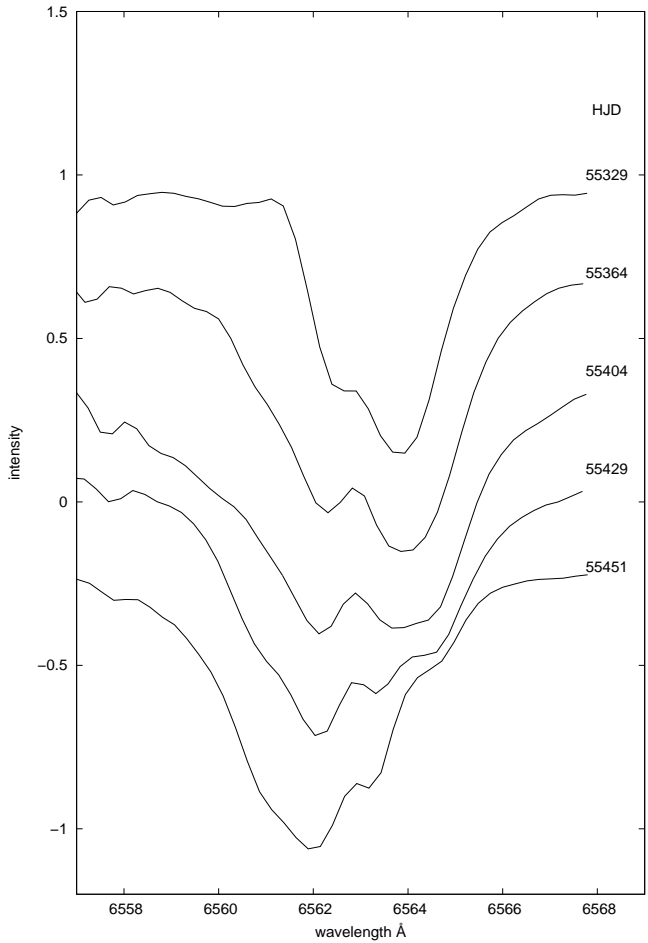}}
\caption{Evolution of the observed \ha profile near mideclipse.}
\label{Halfa1}
\end{figure}

\begin{figure}[t]
\centering \resizebox{\hsize}{!} {\includegraphics{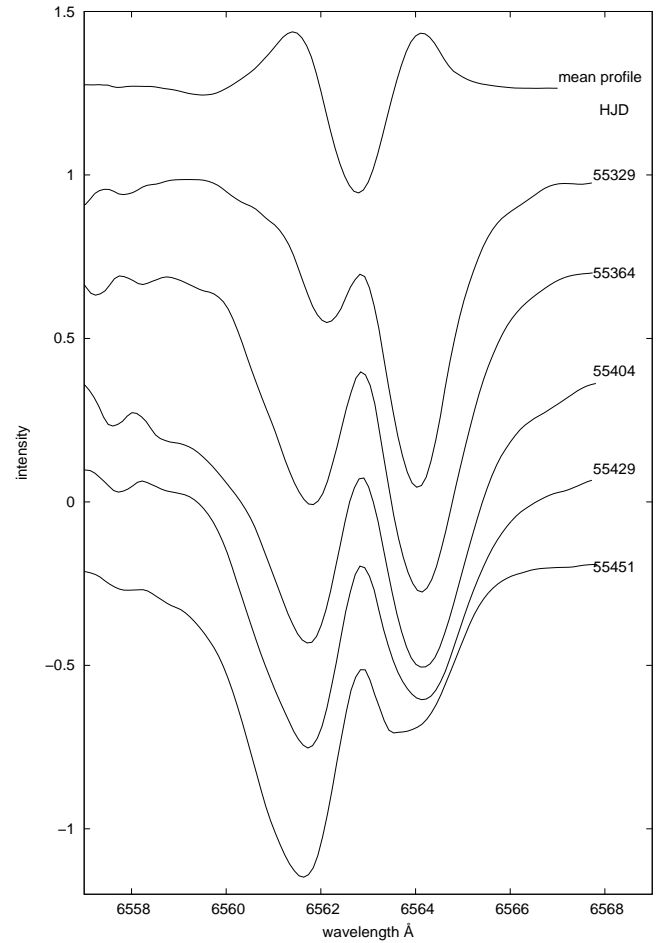}}
\caption{Evolution of the corrected \ha profile, with the mean
out-of-eclipse profile subtracted, near mideclipse.} 
\label{Halfa2}
\end{figure}

To test and explore our hypothesis, we once again used KOREL to
obtain a mean \ha profile, free of telluric lines and intrinsic
variations, using only the spectra observed out of eclipse. This
mean \ha profile was then shifted to the primary's RV, using the
\citet{Chadima2010} orbital solution, and subtracted from the
observed OND spectra. Since the shell absorption should have the
out-of-eclipse \ha as its continuum, we finally renormalized
the subtracted spectra to a continuum value of unity. In
Fig.~\ref{Halfa1}, we show observed \ha line profiles at five
epochs near the date of mideclipse. The complex \ha line profiles
shown here are similar to those reported  by other observers during
the current \ea observing campaign. In Fig.~\ref{Halfa2}, we show
the mean, out-of-eclipse profile and the same five profiles as
in Fig.~\ref{Halfa1}, but with the mean, out-of-eclipse profile
subtracted from them. The result of the subtraction supports our
hypothesis and the validity of our procedure. This is because the
subtracted spectra show simple, double-absorption core profiles
consistent with what is expected from the occulting disk model.
Furthermore, these \ha absorption profiles evolve during the course
of the eclipse in a manner consistent with what is expected from
the occulting disk model, in contrast to the complex observed \ha
profiles of Fig.~\ref{Halfa1}. The red absorption core is much deeper
on the earliest spectrum, when most of the light from the primary
is absorbed by a receding part of the rotating disk. Furthermore,
the ratio of the central intensities of the red/blue absorption cores
gradually decreases through unity near mideclipse. At mideclipse
(the third spectrum in both figures), the two absorption cores are
almost symmetric (after subtraction of the mean, out-of-eclipse profile).

\begin{figure}[t]
\centering \resizebox{\hsize}{!} {\includegraphics{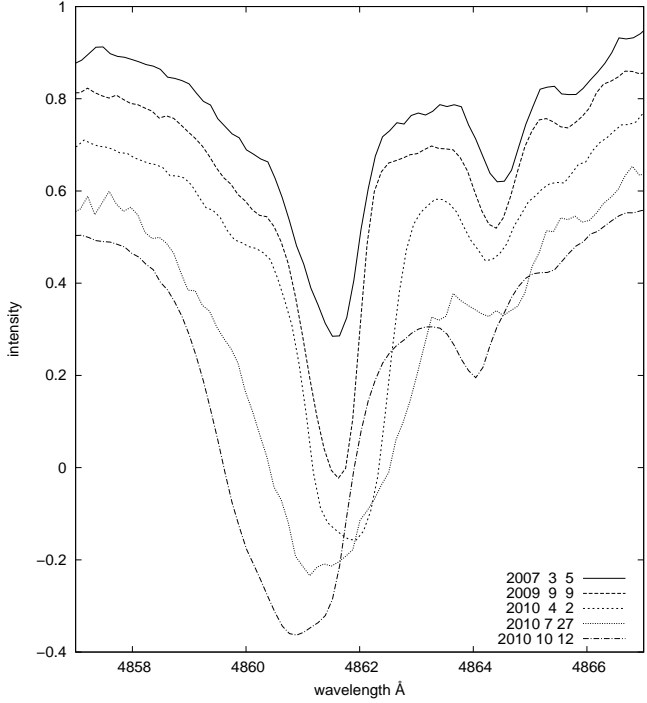}}
\caption{Time evolution of the original \hb profile during the
2009--2011 eclipse.}\label{Hbeta}
\end{figure}

Similar profile changes should also be seen for the \hb line.
We checked 30 available OND spectra covering the \hb region,
observed between 2007 and 2010. Regrettably, we have no out-of-eclipse
\hb spectra and cannot repeat the procedure carried out for \hax.
We therefore only present a selection of five original spectra
in Fig.~\ref{Hbeta}. It is seen that their CI is decreasing
and the absorption core is shifting redward as the mideclipse
approaches. During and after the mideclipse, the absorption core
quickly becomes blueshifted. This indicates that the evolution of
the \hb profile near the current mideclipse is indeed similar to
that of the \ha profile.

\begin{figure}[t]
\centering \resizebox{\hsize}{!} {\includegraphics{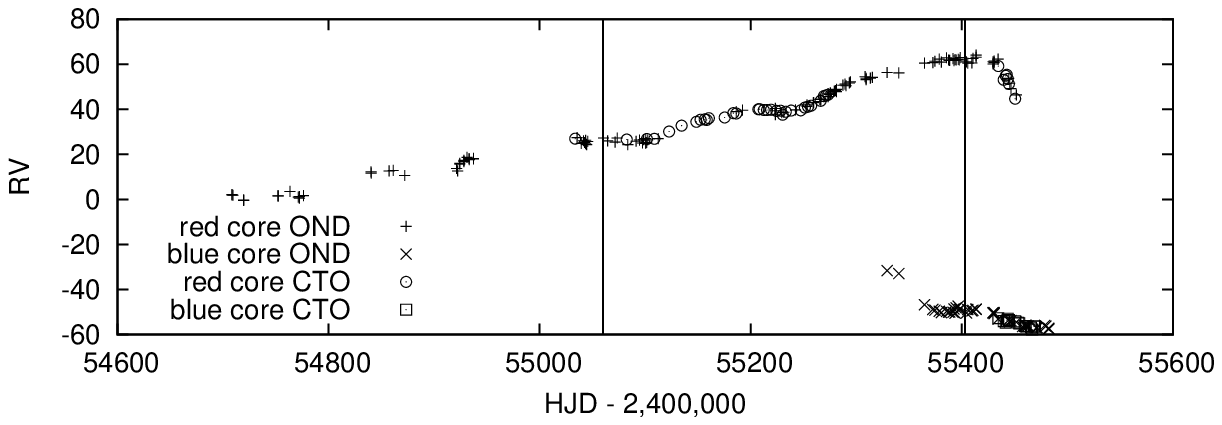}}
\resizebox{\hsize}{!} {\includegraphics{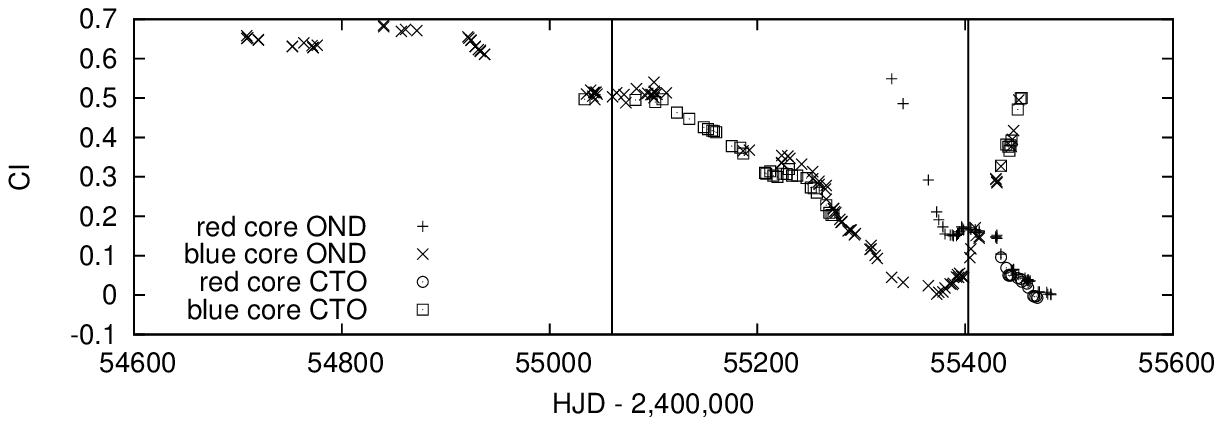}}
\resizebox{\hsize}{!} {\includegraphics{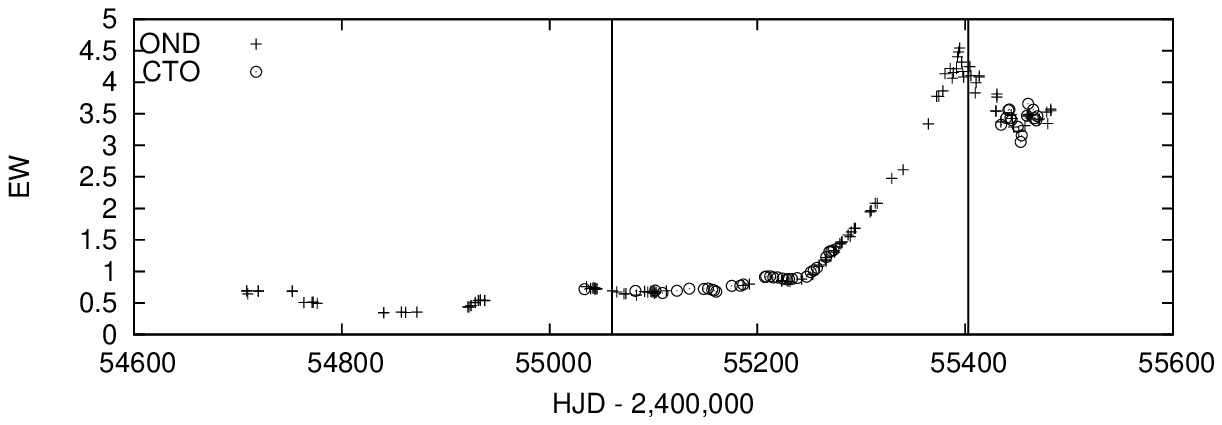}}
\caption{Changes of RV (top), CI (middle), and EW (bottom) of the two
absorption cores of \ha during eclipse, after subtraction of the mean,
out-of-eclipse \ha profile. The first vertical line denotes the
beginning of the current photometric eclipse, the second one indicates
the date of mideclipse.} \label{Ha}
\end{figure}

We measured the RVs and CIs of both \ha cores after the subtraction
of the mean, out-of-eclipse profile, as well as the EWs of the
entire double absorption profile. All these quantities are plotted
in Fig.~\ref{Ha}. It is seen that the CI decreases and the EW increases
with the approach of mideclipse. These changes become more prominent
closer to mideclipse as the light from the primary goes through denser
parts of the occulting disk, resulting in stronger disk absorption. The
RV increases as the line-of-sight absorption through the occulting disk
gradually moves closer to the disk's axis of rotation (and to correspondingly
higher orbital velocities). The blue core, with its RVs in the opposite
sense, first appears 74 days before the mideclipse\footnote{The difference 
between the HJDs of mideclipse and the first obvious appearance of the 
blue core in the spectrum.}, and it deepens quickly, while the the red 
core reaches its maximum strength and gradually weakens thereafter.
The red core finally disappears 49 days after mideclipse, at about the
time when the blue core attains its maximum strength (i.e., minimum CI). 
The interval when the both cores are visible is therefore 123 days. This
is in a good agreement with the estimated time needed for the center of
the companion's occulting disk to transit the disk of the F~star. It is
also seen that, while the EW of the entire \ha absorption line profile 
reaches its maximum several days prior to mideclipse, the red and blue 
absorption cores do not attain equal CI until several days after mideclipse.
One possible explanation of this finding is that the center of the disk
is not perfectly axially symmetric and that there is a denser region
in the receding part of it and a more rarified region in its approaching
part. It is important to continue these \ha observations for the second
half of the current eclipse to confirm the occulting disk model proposed here.

We emphasize that the RVs in Fig.~\ref{Ha} should be corrected for
orbital motion before any analysis. But since the \textquoteleft
shell\textquoteright\  absorption cores originate in the disk around
the secondary, \emph{their RV measurements should be corrected for an
orbital velocity of the secondary, not the primary.} Nevertheless,
uncertainties in present estimates for the mass ratio are too high,
thereby keeping this correction from being inconvincing. However,
it was proven that any reasonable value of the mass radio does not
change the resulting RVs so much that it would affect our conclusions
presented in this section. However, some inconsistencies remain with
the occulting disk scenario.
\begin{enumerate}
\item The average between the first (HJD~2455329) and the last
(HJD~2455452) corrected spectrum for which both cores are visible
is HJD~2455391, which disagrees with the predicted times of the
mideclipse of either HJD~2455413.8\p4.8 \citep{Stefanik2010} or
HJD~2455402.8\p1.0 \citep{Chadima2010}.
\item The RVs of the blue core are gradually {\sl decreasing} during
last few observations, while their increase is expected.
\item The last few CIs of the blue core (of the subtracted spectra) 
are slightly negative, which is, of course, physically impossible.
\end{enumerate}
All of these inconsistencies can be explained if we realize that the
out-of-eclipse \ha profile is highly variable, and it is undoubtedly
also variable during the eclipse, thereby affecting the observed eclipse
profiles. Since we corrected all \ha profiles by subtracting the {\sl mean}
profile, any intrinsic line variations will still affect the individual
profiles.\footnote{Note that \citet{Cha1994} made an unsuccessful
attempt to find periodic behavior in the \ha profile; see also the
first paragraph of Sect.~\ref{per}.} Moreover, the measured quantities
in the absorption profiles could also be partly affected by relatively
strong telluric lines at 6560.555~$\AA$ and 6564.206~$\AA$.

Our hypothesis and results discussed above are potentially good for 
determining the physical properties of the disk (i.e., its density and
velocity structure), which could help in understanding its nature. But 
for a precise analysis of this kind, one should model the light of the
primary going through the various model disks and compare the results
with the observed \ha profiles. This task is beyond the scope of this,
rather qualitative, analysis.

\begin{figure}[t]
\centering \resizebox{\hsize}{!} {\includegraphics{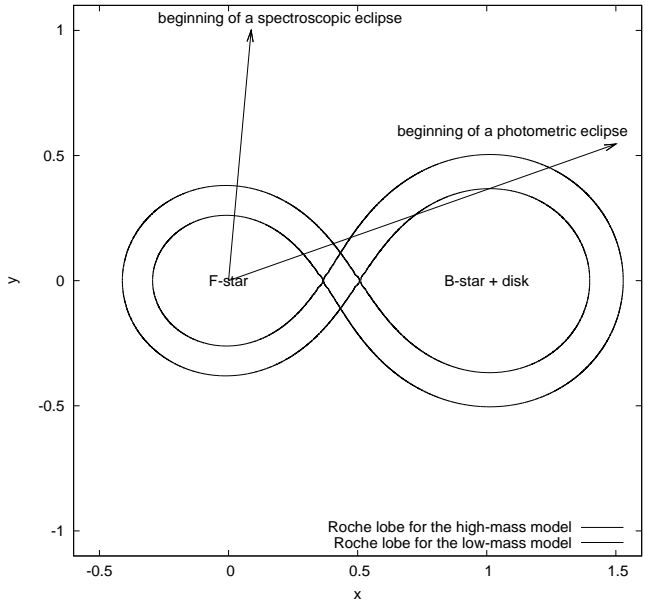}}
\caption{Critical Roche lobes for the high and low-mass models are shown
to scale, with the lines of sight to Earth shown for the beginning of
spectroscopic and photometric eclipse.}
\label{Roche}
\end{figure}

\subsection{Sightlines at the onset of spectroscopic and photometric 
eclipse}\label{eclipses}

As already noted at the beginning of this section, the CI of the \ha
absorption core starts to decrease at least three years before the
predicted beginning of primary eclipse. This observational result can
be used to constrain the structure of the system. We refer to the period
when \ha absorption cores deepen below their out-of-eclipse mean value
as \textquoteleft spectroscopic eclipse\textquoteright. From
Fig.~\ref{Ha_orig}, we estimate that spectroscopic eclipse began
around HJD~2454000. The \citet{Chadima2010} orbital solution
implies that the angle between the sightline at the start of spectroscopic
eclipse and the mideclipse sightline is $\sim$85$^\circ$. JH determined
that the 2009 photometric eclipse began on HJD~2455056 when the sightline
angle was only $\sim$20$^\circ$ relative to the mideclipse line of sight.

In Fig.~\ref{Roche}, we show the critical Roche lobes for the
high- and low-mass models presented in Sect.~\ref{intro}, and the 
sightlines at the onset of spectroscopic and photometric eclipse. 
It is evident that the additional \ha absorption at the start of
spectroscopic eclipse cannot be caused by material near the
secondary because the binary orientation is close to maximum
separation at that time. Therefore there must be some circumbinary
material responsible for this additional absorption, suggestive of
the \citet{Struve1956} model. However, from our observations,
it is seen that this envelope is not homogenous, as proposed by
\citet{Struve1956}.

The line of sight at the onset of photometric eclipse intersects
both critical Roche lobes around the secondary. However, in the case
of the high-mass model, the disk around the secondary would almost
extend to the critical Roche lobe.

\begin{figure}[t]
\centering \resizebox{\hsize}{!} {\includegraphics{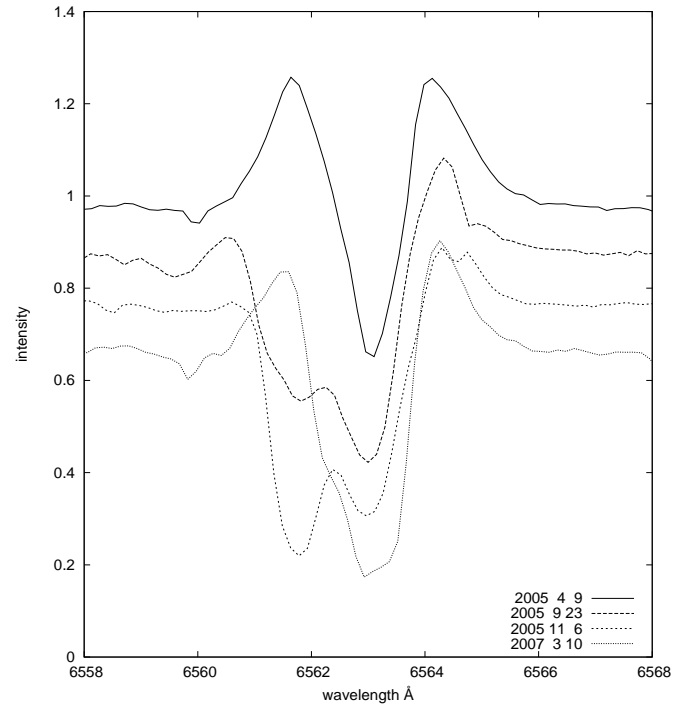}}
\caption{Dramatic \ha profile variation: the sudden appearance
of a deep, blueshifted absorption core, as observed at the DAO during
2005--2006.}
\label{outburst}
\end{figure}

\subsection{A remarkable out-of-eclipse \ha profile variation 
in 2005--2006}\label{event}
An unusual and prominent change in the \ha line profile was observed
during 2005--2006, long before the onset of spectroscopic eclipse. 
The blue emission wing disappeared and was replaced by a deep, 
blueshifted absorption core (Fig.~\ref{outburst}, also Fig.~\ref{ha_fig1}).
Initially, the first spectrum (in Apr 2005) has a \textquoteleft
standard\textquoteright\  \ha profile, while the second and the third
spectra observed later that year show the gradual appearance of the 
blueshifted absorption core. By Mar 2007, the final spectrum of this
series shows a return to the standard \ha profile, but now with
an increase in EW due to the onset of spectroscopic eclipse. Our
observational coverage of this period is not sampled well enough
to permit a reliable estimate of the timescale of the event. However,
considering the first and the last spectrum for which a deviation
from the standard \ha profile was seen, we estimate that the event
lasted from August 2005 to March 2006. A similar behavior of the
\ha profile was also observed and reported by \citet{Schanne2007}.
However, a similar profile variation is not observed for other
spectral lines.

The unusual change in the \ha profile was most probably caused by
additional strong absorption from gaseous material with at an RV
of about --40~\kms relative to the primary. This absorption feature
indicates the presence of either a transient outflow of material 
from the primary star or of some localized circumstellar matter 
producing an \textquoteleft eclipse\textquoteright\  of the primary.

\begin{figure}[t]
\centering \resizebox{\hsize}{!} {\includegraphics{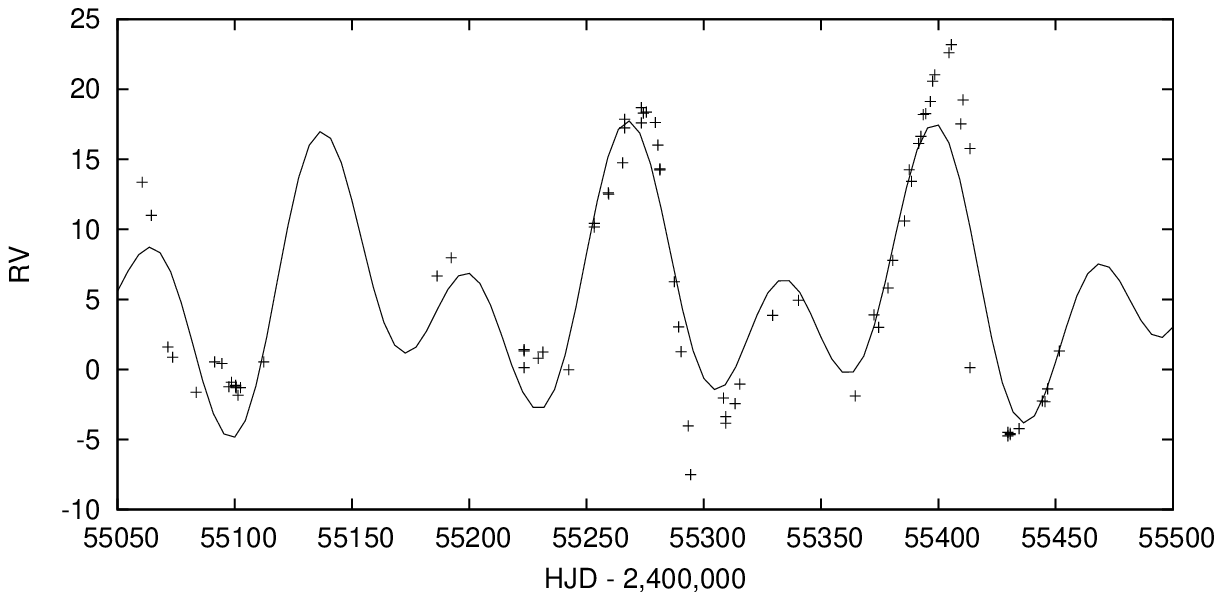}}
\resizebox{\hsize}{!} {\includegraphics{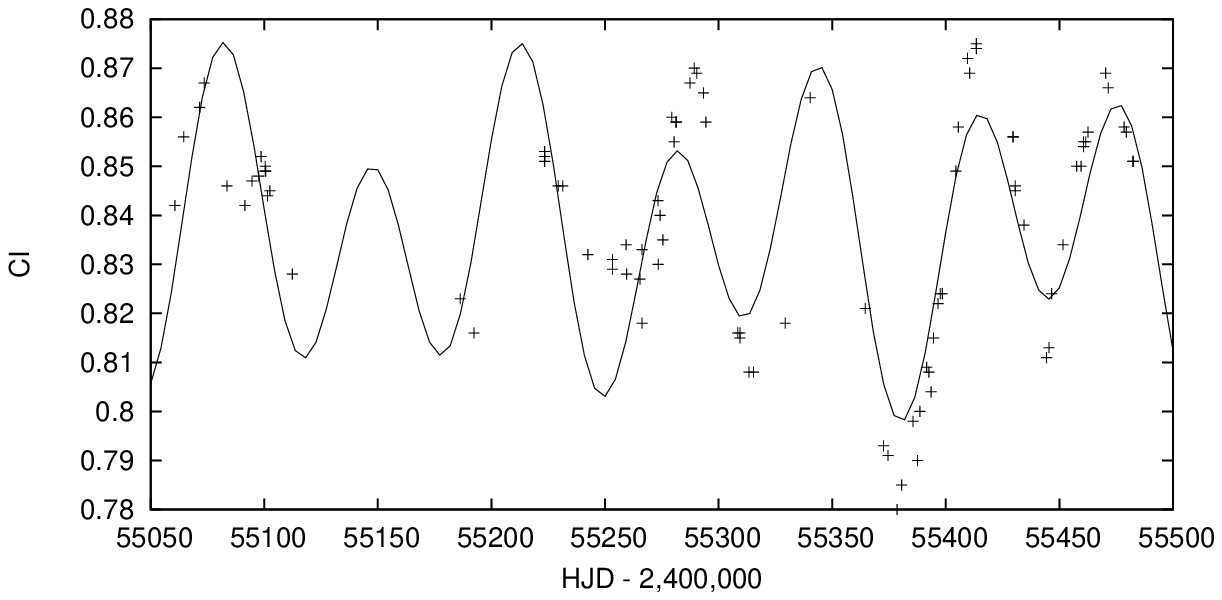}}
\resizebox{\hsize}{!} {\includegraphics{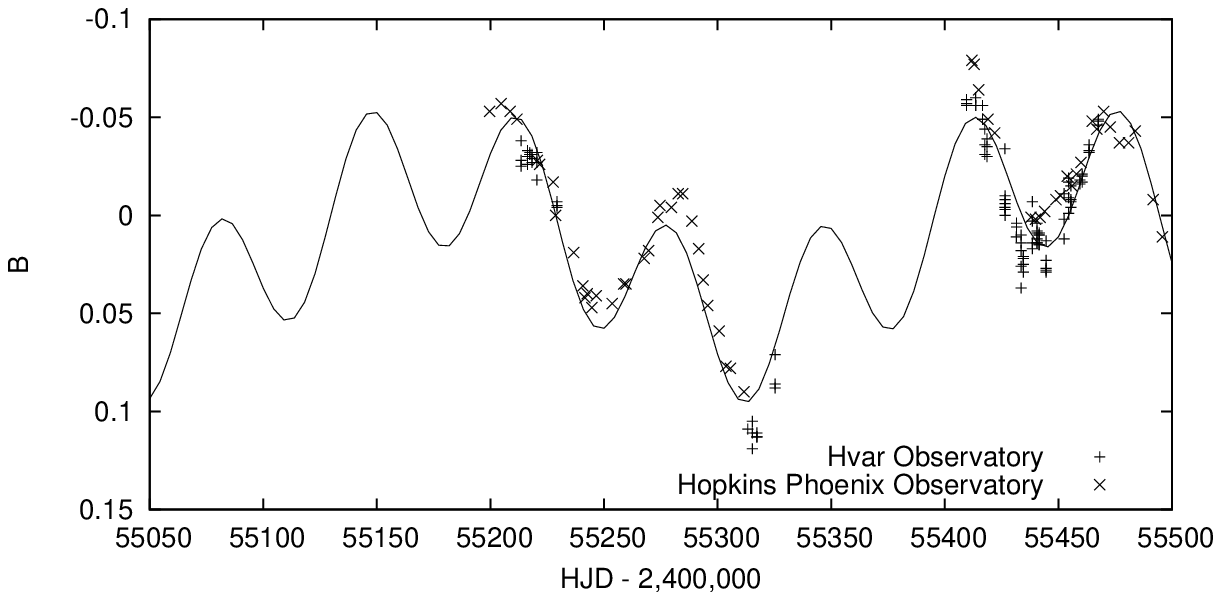}} \caption{RVs
of the \fe lines (upper panel), the CIs of the 6416.919~$\AA$ \fe
line (middle panel) and the B photometric filter (lower panel)
fitted with two sinusoids. In each case the dominant period is 66.21
days, while the secondary period differs among the three cases.}
\label{periods}
\end{figure}

\section{A period analysis of the spectroscopic quantities and \ubv\ 
photometry during the 2009--2011 eclipse}\label{per}
The profiles of the selected spectral lines, measured by us, are all
variable on much shorter timescales than for the binary orbital period.
To find out whether this variability is irregular or (multi)periodic,
we carried out period analyses for individual observables, using two
different computer programs. The first program was the HEC27\footnote{The
program with brief instructions how to use it is available at 
{\tt http://astro.troja.mff.cuni.cz/ftp/hec/HEC27\,.}}, a code 
developed by P.~Harmanec and based on phase dispersion minimalization
\citep{stellingwerf1978}. The second program, PERIOD04, developed by
P.~Lenz and M.~Breger \citep{Lenz2005}, uses Fourier analysis methods;
i.e., the time series of observational data is fitted using a sum
of cosine functions.

Initially, we analyzed the periodic behavior of the \ha profile out
of eclipse. For analysis of the \ha absorption profile RVs and CIs, we
selected the 84 DAO observations prior to HJD~2454913. Later observations 
were excluded because these were affected by the peculiar line profile 
behavior discussed in Sect.~\ref{event} and, subsequently, by the onset 
of spectroscopic eclipse discussed in Sect.~\ref{eclipses}. All out-of-eclipse
measurements were included for analyzing of the emission wings: the RVs,
the CIs of the red and the blue emission, the emission-peak ratio $V/R$,
and the emission strength $(V+R)/2$ where $V$ is the peak intensity of
the blue wing and $R$ is the peak intensity of the red wing. Both the
HEC27 and PERIOD04 analyses found no evidence of periodic behavior
in any of the observables analyzed, a result in accordance with that
of \citet{Cha1994}. We conclude that the physical changes in the
circumstellar matter responsible for the \ha emission are not periodic.

Next, we analyzed the behavior of the RVs and the CIs of the two measured
\fe and \si lines. Since we assumed that the eclipse could affect these
lines and since we did not have enough observations during the eclipse,
we initially analyzed only the out-of-eclipse data. We used both methods
but we also did not find any periodic behavior in the changes of these lines.

\setcounter{table}{3}
\begin{table}
\caption{Periods found during the 2009--2011 \ea eclipse.}
\label{tab4}
\begin{center}
\begin{tabular}{lccc}
\hline\hline\noalign{\smallskip}
 Observable & Period & Phase shift\\
 & (d) &\\
\noalign{\smallskip}\hline\noalign{\smallskip}
\fe and \si RVs & 66.21 & 0.000\\
                & 123.3 & 0.668\\
(just for \si)  & 316.9 & 0.182\\
\noalign{\smallskip}\hline\noalign{\smallskip}
\fe CIs & 66.21 & 0.808\\
        & 119.9 & 0.659\\
\noalign{\smallskip}\hline\noalign{\smallskip}
\si CIs & 66.21 & 0.808\\
        & 268.2 & 0.356\\
\noalign{\smallskip}\hline\noalign{\smallskip}
\ubv\ photometry & 66.21 & 0.323\\
                 & 269.8 & 0.081\\
\noalign{\smallskip}\hline
\end{tabular}
\end{center}
\end{table}

After collecting sufficient observations of the current (2009--2011)
eclipse, we repeated the periodic analyses just for the eclipse period,
motivated by indications of periodic behavior during eclipse. We used
the PERIOD04 code for this analysis, and we found that all the observables
could be reasonably modeled by two sinusoids, except for the RVs of
the \si lines, which needed three harmonic functions to adequately
describe their variations. It is interesting to note that the main
period of $\sim$66 days was common to all the analyzed observables.
In contrast, the secondary period differed among the observables.
However, subsets of the observables shared common secondary periods:
all the RVs had the same secondary periods, as did each of the \fe
line CIs, and each of the \si line CIs (Table~\ref{tab4}). As a final
exercise, we tried to extrapolate these periods to the out-of-eclipse
observations (which have much sparser time coverage), but this failed
to accurately describe the observed variability. There is evidence
that the photometric variability of \ea is quite coherent at some times,
but this coherence is lost on longer timescales. This \textquoteleft
unpredictable\textquoteright\  behavior is presumably due to some
intrinsic variability of the F-type primary star.

We decided to investigate whether similar periodic behavior could
also be found in the photometric data at our disposal. A correction
of the photometry for the eclipse light curve cannot be carried out
reliably since there is no accurate model light curve so far. Using
an improper model could result in the detection of spurious periodicities.
Therefore we only used the photometric observations only from the
eclipse period after the second photometric contact. The third contact
had not yet occurred at the time of writing. We further assumed that
the {\sl only} brightness variations of \ea in the \ubv\ passbands
were due to the intrinsic variability of the only partially eclipsed
F~star \citep{Kloppenborg2010}, i.e., we assumed the eclipse imposed
no further photometric variability during this period. With these
assumptions, we only needed to shift the two datasets of \ubv\ photometry
to a common zero point. We computed average values for both datasets
in all three passbands and subtracted them from the observed values.
Analysis of these data by PERIOD04 revealed that the observations in
all filters can also be described by two harmonic functions with both
periods nearly identical for all filters. The main period was again
$\sim$66 days.

There are small differences in the main and secondary periods found
by this analysis, but these can be interpreted as a consequence of
fitting data with finite noise on a finite time interval. Considering
that the differences in all periods imply phase differences of only a
few percent of a cycle at worst, we simply averaged the individual
periods and their respective phase shifts. These mean quantities were 
subsequently held fixedm, while the fit to the observational data was
repeated in order to determine their amplitudes. The resulting periods 
and phase shifts (with respect to the phase of the main RV period) are
summarized in Table~\ref{tab4}.

The main period was found to be 66.21 days, and it is common to all
analyzed quantities, which proves that it is real. There is a phase
shift of 0.808 between the RVs and the CIs and 0.323 between the RVs
and the \ubv\ photometry. The secondary periods differ among the
various observables, but as already mentioned, related observables
share common secondary periods, which also suggests these periods
are probably real. Figure~\ref{periods} illustrates the final period
fit for the RVs of the \fe lines, for the CIs of the 6416.919~$\AA$
\fe line, and for the $B$ photometric passband.

\section{Conclusions}\label{concl}
Our spectroscopic and photometric analysis of the \ea led to the 
following conclusions as listed below. Since this study does not
include observations from the end of the current eclipse and from
post-eclipse phases, we intend to update this analysis once these
phases have been covered by data from our ongoing observations.
\begin{enumerate}
\item We attempted to disentangle the \ea spectra and recover the
spectra of both components, taking the unknown mass ratio $q=M_1/M_2$
as a free parameter, but this effort was unsuccessful. We conclude
that there is no secondary spectrum hidden in the prominent primary
spectrum. However, it is possible that a secondary spectrum could 
remain hidden if the \ea companion was itself a binary system.
\item We derived a mean out-of-eclipse spectrum of the primary
and compared it with a grid of synthetic spectra for `classical' 
F~stars, using plane-parallel, LTE ATLAS model atmospheres. These
analyses indicated that the F~star is a low-gravity object, but a
good spectral match was not found. However, this analysis should 
be repeated using more appropriate NLTE, spherically symmetric 
model atmospheres.
\item We interpreted the complex behavior of the \ha line during
eclipse as arising from additional absorption in an extended 
\textquoteleft atmosphere\textquoteright\  around the companion's
disk. We subtracted the mean out-of-eclipse \ha line profile from
the profiles observed during eclipse and obtained symmetric
two-component absorption profiles. The behavior of these line
profiles during the course of the eclipse agreed qualitatively
with what is expected from the model presented by \citet{Huang1965}.
This observational result, when combined with modeling of synthetic
spectra for various eclipse phases, should serve to constrain the
density and velocity structure of the companion's disk.
\item A systematic decrease in the central intensity of the \ha line
(\textquoteleft spectroscopic eclipse\textquoteright\  ) began
$\sim$3 years before photometric first contact, owing additional
circumstellar absorption. At the onset of spectroscopic eclipse,
the line of sight to the F star lies far (at an angle of $\sim$85$^\circ$)
from the mideclipse sightline. This implies that circumstellar material
associated with the secondary must extend well beyond the occulting
disk responsible for the photometric eclipses. In contrast, the sightline
at the start of photometric eclipse (first contact) lies only
$\sim$20$^\circ$ away from the mideclipse line of sight.
\item We searched for periodic behavior in the variable, out-of-eclipse 
\ha line, but did not find any periodicity. This implies that the
distribution of circumstellar material around the primary (which may
be a source of the \ha emission) is complex.
\item During eclipse, both the \si and the \fe lines we examined
showed periodic variations in their radial velocities and central 
intensities. The \ubv\ photometry showed a similar periodicity. 
Attempts to extrapolate this periodic solution to the out-of-eclipse 
observations failed. The main period of the eclipse variability
was found to be the same (66.21 days) for the radial velocity,
central intensity, and photometric variations. However, these
observables did not vary in phase, but exhibited different phase
shifts among the variables. The lack of a periodic solution to the 
out-of-eclipse data appears to rule out regular stellar pulsations
as the cause of this variability. The interpretation of the
out-of-eclipse variability remains uncertain.
\end{enumerate}

\begin{acknowledgements}
We would like to thank our colleagues who obtained a few Ond\v{r}ejov
spectra used in this study: Drs.~A.~Kawka,  D.~Kor\v{c}\'akov\'a,
P.~Mayer, and P.~Zasche, and students B.~Ku\v{c}erov\'a and J.~Polster.
We acknowledge the use of the public version of the program KOREL written
by Dr.~P.~Hadrava. We profited from the bibliography maintained by the
NASA/ADS system and the CDS in Strasbourg. The Czech authors were supported
by the grants 205/06/0304, 205/08/H005, 205/09/P476, and P209/10/0715 of
the Czech Science Foundation and also from the research programs
AV0Z10030501 and MSM0021620860. The University of Denver participants
are grateful for support under US NSF grant 10-16678 and the bequest of
William Hershel Womble in support of astronomy at the University of Denver.
\end{acknowledgements}

\bibliographystyle{aa}
\bibliography{citace}

\end{document}